\documentclass[letterpaper,12pt]{article}
\usepackage[letterpaper,total={6.5in, 10in}]{geometry}
\usepackage{cite}
\usepackage{xcolor}
\usepackage{graphicx} 
\usepackage{amsmath}
\usepackage{mathrsfs}
\usepackage{amsfonts}
\usepackage{hyperref}
\hypersetup{
    colorlinks=true,
    linkcolor=blue,
    filecolor=magenta,      
    urlcolor=cyan
}
\allowdisplaybreaks

\title{\bf TURBO CODED OFDM-OQAM USING HILBERT TRANSFORM}
\author{Kasturi Vasudevan$^1$, Surendra Kota$^1$, Lov Kumar$^1$,\\ Himanshu Bhusan
        Mishra$^2$\\
        $^1$Dept of EE IIT Kanpur India;\\
        $^2$Dept of Electronics IIT-ISM Dhanbad Indiaoffset,
        average SNR per bit, BER, matched filter, Hilbert transform,\\
        turbo code.}
\date{}

\begin{document}

\maketitle
\thispagestyle{empty}

\begin{center}
    \bf ABSTRACT
\end{center}
Orthogonal frequency division multiplexing (OFDM) with offset quadrature amplitude
modulation (OQAM) has been widely discussed in the literature and is considered a 
popular waveform for 5th generation (5G) wireless telecommunications and beyond. In 
this work, we show that OFDM-OQAM can be generated using the Hilbert transform and is 
equivalent to single sideband modulation (SSB), that has roots in analog 
telecommunications. The transmit filter for OFDM-OQAM is complex valued whose
real part is given by the pulse corresponding to the root raised cosine spectrum
and the imaginary part is the Hilbert transform of the real part.
The real-valued digital information (message) are passed through the transmit filter
and frequency division multiplexed on orthogonal subcarriers.
The message bandwidth corresponding to each subcarrier is assumed to be narrow enough 
so that the channel can be considered ideal. Therefore, at the receiver, a matched 
filter can used to recover the message. Turbo coding is used to achieve bit-error-rate 
(BER) as low as $10^{-5}$ at an average signal-to-noise ratio (SNR) per bit close to 0 
db. The system has been simulated in discrete time.
\section*{\centering\normalsize\bf INTRODUCTION}
\label{Sec:Intro}
Orthogonal frequency division multiplexing (OFDM)
\cite{6663392,Vasudevan2015,Vasu_Adv_Tele_2017} and
OFDM offset quadrature amplitude modulation (OFDM-OQAM) \cite{5753092}
are the preferred modulation techniques for transmission of
digital information over frequency selective channels, both fading and non-fading.
The variants of OFDM-OQAM
\cite{6852083,7744812,8014377,9339835,9861227}
are known as filter bank multicarrier (FBMC)
\cite{5753092,6852083,7744812,8014377} and
universal filtered multicarrier (UFMC) \cite{7315054,8067703} and
generalized frequency division multiplexing (GFDM)
\cite{7897320,8542819,8968732}
in the literature.

One of the key advantages of OFDM-OQAM/FBMC/UFMC over OFDM is its
immunity against carrier frequency offsets (CFO). In other words, it may not be
necessary for 
an OFDM-OQAM/FBMC/UFMC system  to estimate and cancel the CFO, unlike OFDM.
However, it has been shown in \cite{6663392,Vasudevan2015,Vasu_Adv_Tele_2017} that
it is possible to estimate and cancel the CFO very accurately with large scope for
parallel processing.
The other important feature of OFDM-OQAM is the spectral containment of each subcarrier 
using a transmit filter, which is absent in OFDM. However, OFDM is more attractive than 
OFDM-OQAM in terms of implementation simplicity.

In this work, we demonstrate that OFDM-OQAM/FBMC/UFMC
can be efficiently implemented
using the Hilbert transform at the transmitter and a matched filter at the receiver
\cite{KV_WCAM2023}.
It must be noted that using OQAM would improve the symbol density in time-frequency
space, at the cost of introducing intersymbol interference (ISI) at the matched filter
output and increasing the receiver complexity. Therefore, in this work, we do not use
OQAM and yet have a symbol density in time frequency space greater than unity. We
however retain the nomenclature ``OFDM-OQAM/FBMC/UFMC'' since this work deals with
multicarrier communications.

We use the following notation. Complex quantities are denoted by a tilde e.g.,
$\tilde{p}(t)$, estimates are denoted by hat e.g., $\hat{b}$ and convolution is
denoted by star e.g., $p(t)\star g(t)$. This article is organized as follows.
We first
discuss the theory of OFDM-OQAM, followed by derivation of Hilbert transform of the
pulse corresponding to root-raised cosine spectrum. The discrete-time implementation
issues are discussed next, motivating the need for a modified Hilbert transform.
The discrete-time system model used for computer simulations is presented, followed
by results and conclusions.
\section*{\centering\normalsize\bf THEORY}
\label{Sec:Basic_Concept}
\begin{figure}[tbhp]
\begin{center}
\input{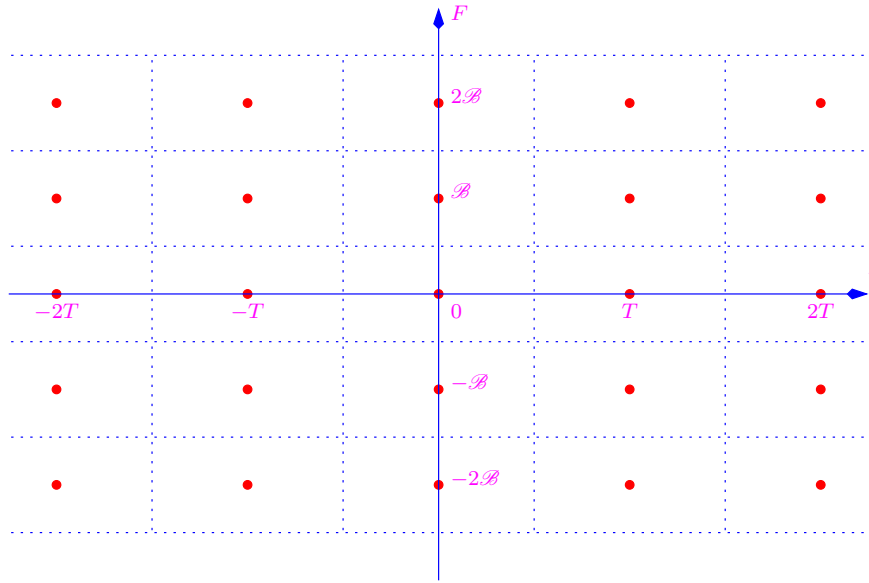}
\caption{Symbol density of multicarrier communication system
         in time-frequency space.}
\label{Fig:FLOGEN2023_OFDM_SD}
\end{center}
\end{figure}

\begin{figure}[tbhp]
\begin{center}
\input{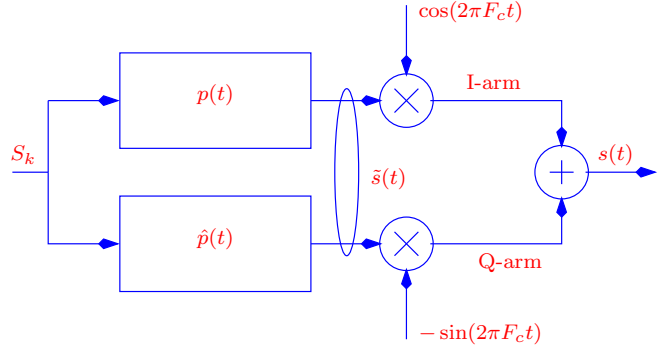}
\caption{Proposed OFDM-OQAM transmitter for each subcarrier.}
\label{Fig:FLOGEN_SSB}
\end{center}
\end{figure}

At the outset, we note that multicarrier communication is equivalent to
frequency division multiplexing.
The time-frequency representation of a multicarrier communication system is
shown in Figure~\ref{Fig:FLOGEN2023_OFDM_SD} \cite{5753092}. The symbol-rate
is $1/T$ baud
and the subcarrier spacing is $\mathscr{B}$. The red colour dots denote symbols.
The subcarrier frequency can be positive or negative, as will be apparent
later.
The complex envelope of a linearly modulated digital signal is given by
\cite{Vasu_Book10}
\begin{align}
\label{Eq:FLOGEN_Eq1}
\tilde{s}(t) & =   \sum_{k=-\infty}^{\infty}
                    S_k \, \tilde{p}(t-kT) \nonumber  \\
             & =    s_I(t) + \mathrm{j}\, s_Q(t) \qquad \mbox{(say)}
\end{align}
where $S_k$ denotes complex-valued symbols drawn from an $M$-ary constellation,
$\tilde{p}(t)$ denotes the (possibly complex-valued) impulse response of
the transmit filter, the
subscripts ``$I,\, Q$'' denote in-phase (real) and quadrature (imaginary)
components respectively.
When the symbols $S_k$ are uncorrelated, the power spectral density of
$\tilde{s}(t)$ in (\ref{Eq:FLOGEN_Eq1}) is \cite{Vasu_Book10}
\begin{equation}
\label{Eq:FLOGEN_Eq1_1}
S_{\tilde{s}}(F) = \frac{P_{\mathrm{av}}}{2T}
                   \left|
                   \tilde{P}(F)
                   \right|^2
\end{equation}
where $\tilde{P}(F)$ is the Fourier transform of $\tilde{p}(t)$ and
$P_{\mathrm{av}}$ denotes the average power of the $M$-ary constellation.
If the transmit filter $\tilde{p}(t)=p(t)$ is real-valued \cite{Vasu_Book10}
having
a root raised cosine (RRC) spectrum with roll-off factor $\rho$, $0<\rho \le 1$,
then the minimum spacing between subcarriers for no aliasing would be
\begin{equation}
\label{Eq:FLOGEN_Eq1_2}
\mathscr{B} = \frac{2(1+\rho)}{2T}
\end{equation}
which is essentially the two-sided bandwidth of $\Tilde{P}(F)$. Therefore, the
symbol density in time-frequency space is the inverse of the area of each
rectangle in Figure~\ref{Fig:FLOGEN2023_OFDM_SD} \cite{5753092}:
\begin{equation}
\label{Eq:FLOGEN_Eq2}
\frac{1}{\mathscr{B}T} = \frac{1}{(1+\rho)} < 1.
\end{equation}
In this article, we propose a complex-valued transmit filter given by
\begin{equation}
\label{Eq:FLOGEN_Eq3}
\tilde{p}(t) = p(t) + \mathrm{j}\, \hat{p}(t)
\end{equation}
where $p(t)$ has an RRC spectrum and $\hat{p}(t)$ is the Hilbert transform of
$p(t)$. The Fourier transform of $\tilde{p}(t)$ in (\ref{Eq:FLOGEN_Eq3}) is
\begin{align}
\label{Eq:FLOGEN_Eq4}
\tilde{P}(F) & =  P(F) +
                 \mathrm{j}\, (-\mathrm{j}\mathrm{sgn}(F)) P(F) \nonumber  \\
             & = \left
                 \{
                 \begin{array}{ll}
                  2P(F)    &  \mbox{for $F>0$}\\
                   P(0)    &  \mbox{for $F=0$}\\
                   0       &  \mbox{for $F<0$}
                 \end{array}
                 \right.
\end{align}
where $P(F)$ is the Fourier transform of $p(t)$ and $\mathrm{sgn}(\cdot)$ is the
signum function \cite{Haykin83,Vasu_AC_PS}. The minimum spacing between subcarriers
in this case is
\begin{equation}
\label{Eq:FLOGEN_Eq5}
\mathscr{B} = \frac{1+\rho}{2T}
\end{equation}
resulting in symbol density in time-frequency space equal to
\begin{equation}
\label{Eq:FLOGEN_Eq6}
\frac{1}{\mathscr{B}T} = \frac{2}{(1+\rho)} \ge 1.
\end{equation}
It is emphasized here that when $\tilde{p}(t)$ is given by (\ref{Eq:FLOGEN_Eq3}),
the symbols $S_k$ have to be real-valued. The reason is as follows. The complex
envelope in (\ref{Eq:FLOGEN_Eq1}) can be written as
\begin{equation}
\label{Eq:FLOGEN_Eq7}
\tilde{s}(t) = \tilde{s}_1(t) \star \tilde{p}(t)
\end{equation}
where ``$\star$'' denotes convolution and
\begin{align}
\label{Eq:FLOGEN_Eq8}
\tilde{s}_1(t) & =   \sum_{k=-\infty}^{\infty}
                      S_k \, \delta_D(t-kT) \nonumber  \\
               & =    s_{1,\, I}(t) + \mathrm{j}\, s_{1,\, Q}(t)
                     \qquad \mbox{(say)}
\end{align}
where $\delta_D(\cdot)$ denotes the Dirac-delta function
\cite{Haykin83,Vasu_AC_PS}. At the receiver,
we use a matched filter given by
$\tilde{p}^*(-t)$. The matched filter output would be given by \cite{Vasu_Book10}
\begin{equation}
\label{Eq:FLOGEN_Eq9}
\tilde{y}(t) = \tilde{s}_1(t) \star \tilde{p}(t) \star \tilde{p}^*(-t).
\end{equation}
Now
\begin{align}
\label{Eq:FLOGEN_Eq10}
\tilde{p}(t)
\star
\tilde{p}^*(-t) & = \left[
                     p(t) + \mathrm{j}\, \hat{p}(t) 
                    \right]
                    \star
                    \left[
                     p(-t) - \mathrm{j}\, \hat{p}(-t) 
                    \right]                              \nonumber  \\
                & =  p(t) \star p(-t) +
                    \hat{p}(t) \star \hat{p}(-t) +
                    \mathrm{j}\,
                    \left[
                    \hat{p}(t) \star p(-t) -
                     p(t) \star \hat{p}(-t)
                    \right].
\end{align}
It can be shown that for real-valued $p(t)$ \cite{Haykin83,Vasu_AC_PS}
\begin{align}
\label{Eq:FLOGEN_Eq11}
\hat{p}(t)
\star
\hat{p}(-t) & =  p(t) \star p(-t)                \nonumber  \\
            & =  R_{pp}(t)                       \nonumber  \\
\hat{p}(t)
\star
 p(-t)      & =  R_{\hat{p}p}(t)                 \nonumber  \\
            & = -p(t) \star \hat{p}(-t)          \nonumber  \\
            & = -R_{p\hat{p}}(t).
\end{align}
Substituting (\ref{Eq:FLOGEN_Eq10}) and (\ref{Eq:FLOGEN_Eq11}) in
(\ref{Eq:FLOGEN_Eq9}) we obtain the matched filter output as
\begin{equation}
\label{Eq:FLOGEN_Eq12}
\tilde{y}(t) = \sum_{k=-\infty}^{\infty}
                2S_k
               \left[
                R_{pp}(t-kT) +
               \mathrm{j}\,
                R_{\hat{p}p}(t-kT)
               \right].
\end{equation}
The condition for zero ISI is
\begin{equation}
\label{Eq:FLOGEN_Eq13}
R_{pp}(mT) = \delta_K(mT)
\end{equation}
where $\delta_K(\cdot)$ is the Kronecker delta function \cite{Vasu_Book10}.
However, $R_{\hat{p}p}(t)$ does not satisfy the zero ISI
condition. Hence using (\ref{Eq:FLOGEN_Eq13}), the symbol-rate sampler at the
matched filter output would yield
\begin{align}
\label{Eq:FLOGEN_Eq14}
\tilde{y}(nT) & = \sum_{k=-\infty}^{\infty}
                   2S_k
                  \left(
                   R_{pp}(nT-kT) +
                  \mathrm{j}\,
                   R_{\hat{p}p}(nT-kT)
                  \right)                               \nonumber  \\
              & =  2S_n +
                   2\mathrm{j}\,
                  \sum_{k=-\infty}^{\infty}
                   S_k
                   R_{\hat{p}p}(nT-kT).
\end{align}
It is clear from (\ref{Eq:FLOGEN_Eq14}) that implementing the matched filter as
given in (\ref{Eq:FLOGEN_Eq9}) and (\ref{Eq:FLOGEN_Eq10}) with complex-valued
symbols $S_k$ is not possible, due to crosstalk (interference between in-phase
and quadrature components). Therefore, the symbols $S_k$ have
to be real-valued. One possible implementation of the proposed OFDM-OQAM/FBMC/UFMC
transmitter is shown in Figure~\ref{Fig:FLOGEN_SSB}.
\begin{figure}[tbhp]
\begin{center}
\input{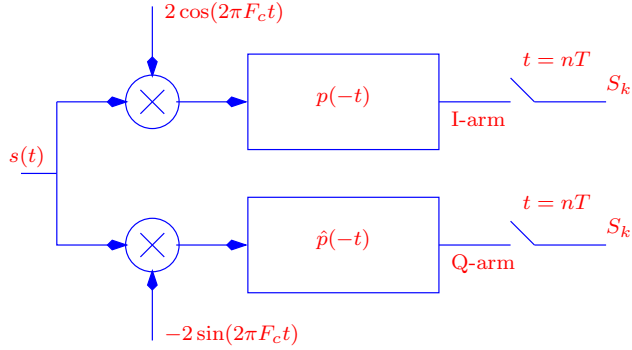}
\caption{Proposed OFDM-OQAM receiver for each subcarrier.}
\label{Fig:FLOGEN_SSB_Rx}
\end{center}
\end{figure}
The corresponding receiver is shown in Figure~\ref{Fig:FLOGEN_SSB_Rx}. In the
next section, we derive $\hat{p}(t)$ when $p(t)$ has an RRC spectrum.
\section*{\centering\normalsize\bf HILBERT TRANSFORM OF RRC SPECTRUM}
\label{Sec:HT_RRC}
The RRC spectrum which is the Fourier transform of $p(t)$ is given by
\cite{Vasu_Book10}
\begin{equation}
\label{Eq:FLOGEN_Eq15}
P(F)        = \left
              \{
              \begin{array}{ll}
              \frac{1}{\sqrt{2B}} & \mbox{for $-F_1 \le F \le F_1$}  \\
              \frac{1}{\sqrt{2B}}
              \cos
              \left(
              \frac{\pi(|F|-F_1)}{4B-4F_1}
              \right)             & \mbox{for $F_1 \le |F| \le 2B-F_1$}  \\
               0                  & \mbox{elsewhere}
              \end{array}
              \right.
\end{equation}
where
\begin{align}
\label{Eq:FLOGEN_Eq16}
2B   & \stackrel{\Delta}{=}   \frac{1}{T}               \nonumber  \\
\rho & \stackrel{\Delta}{=}    1- \frac{F_1}{B}.
\end{align}
Therefore
\begin{equation}
\label{Eq:FLOGEN_Eq17}
\hat{p}(t) = \int_{F=-\infty}^{\infty}
              -\mathrm{j}\,
               \mathrm{sgn}(F) P(F)
               \mathrm{e}^{\mathrm{j}\, 2\pi Ft} \, dF
\end{equation}
where $P(F)$ is given by (\ref{Eq:FLOGEN_Eq15}) and (\ref{Eq:FLOGEN_Eq16}).
We have
\begin{align}
\label{Eq:FLOGEN_Eq18}
I_1 & = \int_{F=-F_1}^{F_1}
        -
        \mathrm{j}\,
        \mathrm{sgn}(F)
        \frac{1}{\sqrt{2B}}
        \mathrm{e}^{\mathrm{j}\, 2\pi Ft} \, dF      \nonumber  \\
    & = \frac{1}{\pi t\sqrt{2B}}
        \left[
         1 - \cos(2\pi F_1 t)
        \right].
\end{align}
Similarly
\begin{align}
\label{Eq:FLOGEN_Eq19}
I_2 & = \int_{F=F_1}^{2B-F_1}
        -
        \mathrm{j}\,
        \frac{1}{\sqrt{2B}}
        \cos
        \left(
        \frac{\pi(F-F_1)}{4B-4F_1}
        \right)
        \mathrm{e}^{\mathrm{j}\, 2\pi Ft} \, dF                \nonumber  \\
    &   \qquad +
        \int_{F=-(2B-F_1)}^{-F_1}
        \mathrm{j}\,
        \frac{1}{\sqrt{2B}}
        \cos
        \left(
        \frac{\pi(-F-F_1)}{4B-4F_1}
        \right)
        \mathrm{e}^{\mathrm{j}\, 2\pi Ft} \, dF                \nonumber  \\
    & = \frac{1}{\sqrt{2B}}
        \int_{F=B(1-\rho)}^{B(1+\rho)}
        \left\{
        \sin
        \left[
         2\pi Ft +
        \frac{\pi(F-F_1)}{4B-4F_1}
        \right] +
        \sin
        \left[
         2\pi Ft -
        \frac{\pi(F-F_1)}{4B-4F_1}
        \right]
        \right\} \, dF                                         \nonumber  \\
    & = \frac{1}{\sqrt{2B}}
        \int_{F=B(1-\rho)}^{B(1+\rho)}
        \left\{
        \sin
        \left[
        \alpha F - \beta
        \right] +
        \sin
        \left[
        \beta - \gamma F
        \right]
        \right\} \, dF                                         \nonumber  \\
    & = \frac{1}{\gamma\sqrt{2B}}
        \left\{
        \cos
        \left[
        \beta  - \gamma B(1+\rho)
        \right] -
        \cos
        \left[
        \beta - \gamma B(1-\rho)
        \right]
        \right\}                                               \nonumber  \\
    &   \qquad -
        \frac{1}{\alpha\sqrt{2B}}
        \left\{
        \cos
        \left[
        \alpha B(1+\rho) - \beta
        \right] -
        \cos
        \left[
        \alpha B(1-\rho) - \beta
        \right]
        \right\}
\end{align}
where
\begin{align}
\label{Eq:FLOGEN_Eq20}
F_1     & =  B(1-\rho)                                \nonumber  \\
2B-F_1  & =  B(1+\rho)                                \nonumber  \\
4B-4F_1 & =  4B\rho                                   \nonumber  \\
\alpha  & = \frac{\pi(1+8Bt\rho)}{4B\rho}             \nonumber  \\
\beta   & = \frac{\pi F_1}{4B\rho}                    \nonumber  \\
\gamma  & = \frac{\pi(1-8Bt\rho)}{4B\rho}.
\end{align}
Now from (\ref{Eq:FLOGEN_Eq20})
\begin{align}
\label{Eq:FLOGEN_Eq21}
\beta - \gamma B & = \frac{-\pi}{4} + 2\pi Bt              \nonumber  \\
\gamma B\rho     & = \frac{\pi}{4} - 2\pi Bt\rho           \nonumber  \\
\alpha B - \beta & = \frac{\pi}{4} + 2\pi Bt               \nonumber  \\
\alpha B\rho     & = \frac{\pi}{4} + 2\pi Bt\rho.
\end{align}
Substituting (\ref{Eq:FLOGEN_Eq21}) in (\ref{Eq:FLOGEN_Eq19}) we obtain
\begin{align}
\label{Eq:FLOGEN_Eq22}
I_2 & = \frac{1}{\sqrt{2B}}
        \sin(2\pi Bt(1+\rho))
        \left\{
        \frac{1}{\gamma} + \frac{1}{\alpha}
        \right\}                                       \nonumber  \\
    &   \qquad -
        \frac{1}{\sqrt{2B}}
        \cos(2\pi Bt(1-\rho))
        \left\{
        \frac{1}{\gamma} - \frac{1}{\alpha}
        \right\}.
\end{align}
Now
\begin{align}
\label{Eq:FLOGEN_Eq23}
\frac{1}{\gamma} -
\frac{1}{\alpha} & = \frac{64B^2\rho^2 t}
                          {\pi(1-64B^2t^2\rho^2)}            \nonumber  \\
\frac{1}{\gamma} +
\frac{1}{\alpha} & = \frac{8B\rho}
                          {\pi(1-64B^2t^2\rho^2)}.
\end{align}
Substituting (\ref{Eq:FLOGEN_Eq23}) in (\ref{Eq:FLOGEN_Eq22}) we get
\begin{align}
\label{Eq:FLOGEN_Eq24}
I_2 & = \frac{1}{\sqrt{2B}}
        \sin(2\pi Bt(1+\rho))
        \left\{
        \frac{8B\rho}
             {\pi(1-64B^2t^2\rho^2)}
        \right\}                                       \nonumber  \\
    &   \qquad -
        \frac{1}{\sqrt{2B}}
        \cos(2\pi Bt(1-\rho))
        \left\{
        \frac{64B^2\rho^2 t}
             {\pi(1-64B^2t^2\rho^2)}
        \right\}.
\end{align}
Finally $\hat{p}(t)$ in (\ref{Eq:FLOGEN_Eq17}) is given by
\begin{equation}
\label{Eq:FLOGEN_Eq25}
\hat{p}(t) = I_1 + I_2
\end{equation}
where $I_1,\, I_2$ are given by (\ref{Eq:FLOGEN_Eq18}) and
(\ref{Eq:FLOGEN_Eq24}) respectively.
\section*{\centering\normalsize\bf DISCRETE TIME IMPLEMENTATION}
\label{Sec:Implement}
Theoretically when $p(t)$ has an RRC spectrum, both $p(t),\, \hat{p}(t)$ have an
infinite time span. In practice they have to be truncated, in which case
(\ref{Eq:FLOGEN_Eq13}) is only approximately valid. Moreover, the transmitter
and receiver in Figures~\ref{Fig:FLOGEN_SSB} and \ref{Fig:FLOGEN_SSB_Rx}
have to be implemented in discrete time. In this section, we explore
these issues.

\begin{figure}[tbhp]
\begin{center}
\input{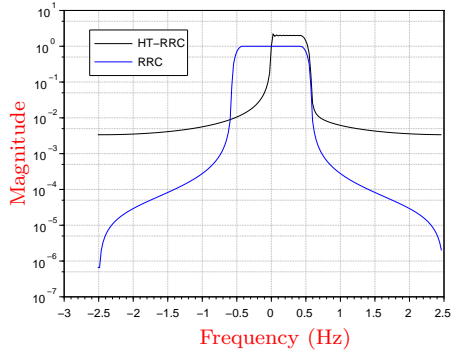}
\caption{Magnitude response of $p(t)$ and $\tilde{p}(t)$.
         $|P(0)|=1$ for RRC, $\left|\tilde{P}(0)\right|=1$ for HT-RRC.}
\label{Fig:HT_RRC_Spec1}
\end{center}
\end{figure}
The magnitude response of $p(mT_s)$ (RRC) and
$\tilde{p}(mT_s)$ (HT-RRC) given by
\begin{equation}
\label{Eq:FLOGEN_Eq25_1}
\tilde{p}(mT_s)=p(mT_s)+\mathrm{j}\,\hat{p}(mT_s)
\end{equation}
is shown in Figure~\ref{Fig:HT_RRC_Spec1} for $T=1$ sec, $\rho=0.161$ and
sampling frequency $F_s=1/T_s=5$ Hz. Both $p(mT_s)$ and $\hat{p}(mT_s)$ lie
in the range $-M\le m \le M$ for $M=100$. The parameter $M$ is referred to as
the one-sided window length. Note that in Figure~\ref{Fig:HT_RRC_Spec1},
$\tilde{P}(\cdot)$ is the discrete Fourier transform of $\tilde{p}(\cdot)$
given in (\ref{Eq:FLOGEN_Eq25_1}).

The ratio of signal-to-interference (SIR) power is defined as
\begin{equation}
\label{Eq:WFLOGEN_Eq26}
\mbox{SIR} =  10
             \log_{10}
             \left[
             \frac{R_{gg}^2(0)}{\sum_{\substack{n\\n\ne 0}}R_{gg}^2(nT)}
             \right] \quad \mbox{db.}
\end{equation}
where $g(\cdot)$ is $p(\cdot)$ or $\hat{p}(\cdot)$. Note that
\begin{align}
\label{Eq:FLOGEN_Eq27}
R_{gg}(mT_s) & =  g(mT_s) \star g(-mT_s)            \nonumber  \\
R_{gg}(nT)   & =  R_{gg}(nIT_s)                     \nonumber  \\
I            & = \frac{T}{T_s}
\end{align}
where ``$\star$'' denotes (discrete-time) linear convolution and $I$ is
the interpolation factor.
\begin{table}[tbhp]
\begin{center}
\caption{SIR of $R_{\hat{p}\hat{p}}(nT)$ and $R_{pp}(nT)$.}
\input{flogen_sir1.pstex_t}
\label{Tbl:FLOGEN2023_SIR1}
\end{center}
\end{table}
The SIR in decibels is shown in Table~\ref{Tbl:FLOGEN2023_SIR1} for different
values of the one-sided window length $M$ when $T=1$ sec, $\rho=0.161$,
$F_s=5=1/T_s$ Hz. We find from Table~\ref{Tbl:FLOGEN2023_SIR1} that the SIR
of $R_{\hat{p}\hat{p}}(t)$ is much lower than $R_{pp}(t)$ for the same values
of $M$. This is probably because the signum function in the Hilbert transform
(see (\ref{Eq:FLOGEN_Eq17})) has a discontinuity at $F=0$. In the next section,
we discuss the modified Hilbert transform that avoids this discontinuity.
\section*{\centering\normalsize\bf MODIFIED HILBERT TRANSFORM}
\label{Sec:Mod_HT}
\begin{figure}[tbhp]
\begin{center}
\input{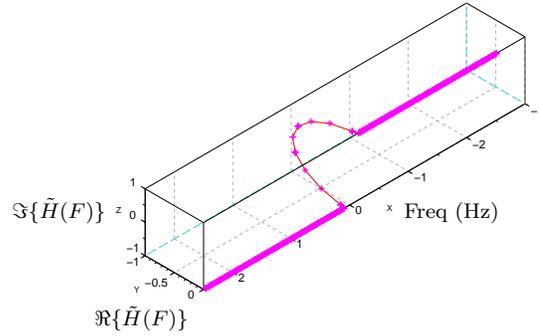}
\caption{Frequency response of the modified Hilbert transformer for
         $a=0.25$, $\rho=0.161$, $T=1$ sec.}
\label{Fig:HT_Spec_Mod}
\end{center}
\end{figure}

Consider the modified Hilbert transform given by \cite{KV_WCAM2023}
\begin{equation}
\label{Eq:FLOGEN_Eq28}
\tilde{H}(F) = \left\{
               \begin{array}{ll}
               \mathrm{e}^{\,\mathrm{j}[\pi(F+aF_1)/(2aF_1)+\pi/2]}  &
               \mbox{for $|F|\le aF_1$} \\
               -
               \mathrm{j}                                            &
               \mbox{for $F\ge aF_1$} \\
               \mathrm{j}                                            &
               \mbox{for $F\le -aF_1$}
               \end{array}
               \right.
\end{equation}
for $0<a\le 1$, which is plotted in Figure~\ref{Fig:HT_Spec_Mod}.
Hence the modified Hilbert transform of $p(t)$ can be written as
\begin{align}
\label{Eq:FLOGEN_Eq29}
\hat{p}_{\mathrm{m}}(t)
             & = \int_{F=-\infty}^{\infty}
                 \tilde{H}(F) P(F)
                 \mathrm{e}^{\,\mathrm{j}\, 2\pi Ft}
                 \, dF                                  \nonumber  \\
             & =  I_{11} + I_{12} + I_2
\end{align}
where $I_2$ is given by (\ref{Eq:FLOGEN_Eq24}) and
\begin{align}
\label{Eq:FLOGEN_Eq30}
I_{11} & = \int_{F=-aF_1}^{aF_1}
           \tilde{H}(F) P(F)
           \mathrm{e}^{\,\mathrm{j}\, 2\pi Ft}
           \, dF                                        \nonumber  \\
       & = \frac{-1}{\sqrt{2B}}
           \int_{F=-aF_1}^{aF_1}
           \mathrm{e}^{\,\mathrm{j}\,\pi F/(2aF_1)}
           \mathrm{e}^{\,\mathrm{j}\, 2\pi Ft}
           \, dF                                        \nonumber  \\
       & = \frac{-2aF_1}{\sqrt{2B}}
           \mbox{sinc}
           \left(
            aF_1A_1
        \right)
\end{align}
where
\begin{align}
\label{Eq:FLOGEN_Eq31}
A_1            & = 2t + \frac{1}{2aF_1}                 \nonumber  \\
\mbox{sinc}(x) & = \frac{\sin(\pi x)}{\pi x}.
\end{align}
Similarly $I_{12}$ in (\ref{Eq:FLOGEN_Eq29}) is equal to
\begin{align}
\label{Eq:FLOGEN_Eq32}
I_{12} & = \frac{-\mathrm{j}}{\sqrt{2B}}
           \int_{aF_1}^{F_1}
           \mathrm{e}^{\,\mathrm{j}2\pi Ft}
           \, dF  +
           \frac{\mathrm{j}}{\sqrt{2B}}
           \int_{-F_1}^{-aF_1}
           \mathrm{e}^{\,\mathrm{j}2\pi Ft}
           \, dF                                         \nonumber  \\
       & = \frac{2F_1(1+a)}{\sqrt{2B}}
           \mbox{sinc}(F_1 t(1+a))
           \sin(\pi F_1 t(1-a)).
\end{align}
\begin{figure}[tbhp]
\begin{center}
\input{ht_rrc_spec3.pstex_t}
\caption{Magnitude response of $p(t)$ and $\tilde{p}_{\mathrm{m}}(t)$.
         $|P(0)|=1$ for RRC, $\left|\tilde{P}_{\mathrm{m}}(0)\right|=1$
         for HT-RRC.}
\label{Fig:HT_RRC_Spec3}
\end{center}
\end{figure}
The magnitude response of $p(mT_s)$ (RRC) and
$\tilde{p}_{\mathrm{m}}(mT_s)$ (mHT-RRC) given by
\begin{equation}
\label{Eq:FLOGEN_Eq33}
\tilde{p}_{\mathrm{m}}(mT_s)=p(mT_s)+\mathrm{j}\,\hat{p}_{\mathrm{m}}(mT_s)
\end{equation}
is shown in Figure~\ref{Fig:HT_RRC_Spec3} for $T=1$ sec, $\rho=0.161$
and sampling frequency $F_s=1/T_s=5$ Hz. Both $p(mT_s)$ and
$\hat{p}_{\mathrm{m}}(mT_s)$ lie in the range $-M\le m \le M$ for $M=100$. Note
that in Figure~\ref{Fig:HT_RRC_Spec3}, $\tilde{P}_{\mathrm{m}}(\cdot)$ is the
discrete Fourier transform of $\tilde{p}_{\mathrm{m}}(\cdot)$ given in
(\ref{Eq:FLOGEN_Eq33}).
\begin{table}[tbhp]
\begin{center}
\caption{SIR of $R_{\hat{p}_{\mathrm{m}}\hat{p}_{\mathrm{m}}}(nT)$
         and $R_{pp}(nT)$.}
\input{flogen_sir2.pstex_t}
\label{Tbl:FLOGEN2023_SIR2}
\end{center}
\end{table}
We find from Table~\ref{Tbl:FLOGEN2023_SIR2} that the SIR of
$R_{\hat{p}_{\mathrm{m}}\hat{p}_{\mathrm{m}}}(nT)$
is comparable to $R_{pp}(nT)$.
\section*{\centering\normalsize\bf DISCRETE TIME SYSTEM MODEL}
\label{Sec:Discrete_Sys_Model}
\begin{figure}[tbhp]
\begin{center}
\input{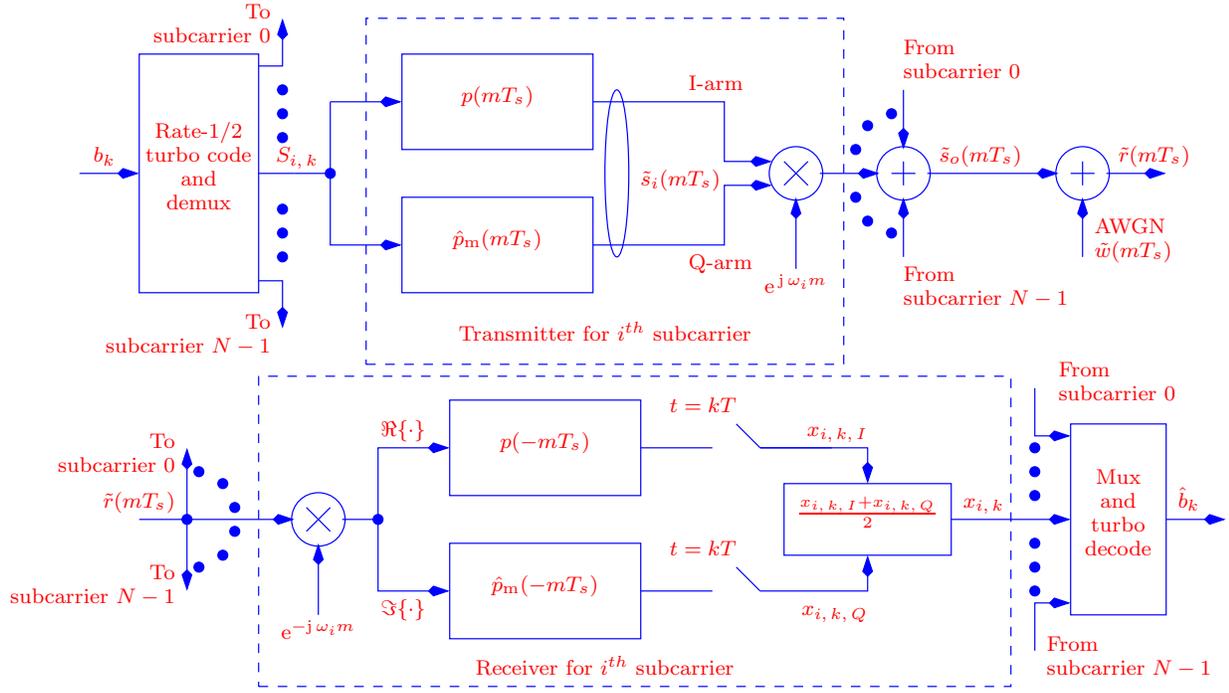}
\caption{Discrete time system model for OFDM-OQAM.}
\label{Fig:FLOGEN_Dis_Sys_Model}
\end{center}
\end{figure}
The discrete time system model used for computer simulations is shown in
Figure~\ref{Fig:FLOGEN_Dis_Sys_Model}. The input $b_k$ is arranged into
frames of length $L_{d1}$ bits and given to the rate-$1/2$ turbo code
\cite{6663392,Vasudevan2015,Vasu_Adv_Tele_2017}. The output of the turbo code
of length $L_d=2L_{d1}$ bits is mapped to binary phase shift keyed (BPSK)
symbols $S_{i,\, k}=\pm 1$, demultiplexed and transmitted simultaneously
over $N$ subcarriers. The frequency of the $i^{th}$ subcarrier is given by
\begin{equation}
\label{Eq:FLOGEN_Eq34}
\omega_i = 2\pi i/N \quad \mbox{radians}   \quad \mbox{for $0\le i \le N-1$}
\end{equation}
where $N=I$ is the total number of subcarriers and $I$ is given by
(\ref{Eq:FLOGEN_Eq27}). The overall OFDM-OQAM signal is given by
\begin{equation}
\label{Eq:FLOGEN_Eq34_1_1}
\tilde{s}_o(mT_s) = \sum_{i=0}^{N-1}
                    \tilde{s}_i(mT_s)
                    \mathrm{e}^{\,\mathrm{j}\, \omega_i m}
\end{equation}
where
\begin{equation}
\label{Eq:FLOGEN_Eq34_1_2}
\tilde{s}_i(mT_s) = \sum_{k}
                     S_{i,\, k}
                    \tilde{p}_{\mathrm{m}}(mT_s - kT)
\end{equation}
where $\tilde{p}_{\mathrm{m}}(\cdot)$ is given by (\ref{Eq:FLOGEN_Eq33}).
\begin{figure}[tbhp]
\begin{center}
\input{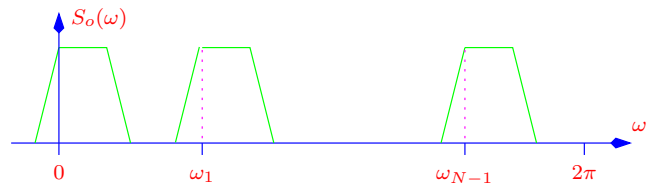}
\caption{Spectrum of the overall OFDM-OQAM signal $\tilde{s}_o(mT_s)$.}
\label{Fig:FLOGEN_Spec_SC}
\end{center}
\end{figure}
The spectrum of the complex-valued overall OFDM-OQAM signal
$\tilde{s}_o(mT_s)$ is shown in Figure~\ref{Fig:FLOGEN_Spec_SC}, where
\begin{equation}
\label{Eq:FLOGEN_Eq34_1}
\omega = 2\pi F T_s \quad \mbox{radians}
\end{equation}
where $F$ is the frequency in Hz.
Recall that the sampling frequency $1/T_s$ Hz maps to $2\pi$ in the digital
frequency domain.
The variance per dimension of complex-valued additive white Gaussian
noise (AWGN) is
\begin{equation}
\label{Eq:FLOGEN_Eq35}
\frac{1}{2}
 E
\left[
\left|
\tilde{w}(mT_s)
\right|^2
\right] = \sigma_w^2.
\end{equation}
The in-phase and quadrature components of $\tilde{w}(mT_s)$ are assumed to be
independent. At the receiver we have for the $i^{th}$ subcarrier
\begin{equation}
\label{Eq:FLOGEN_Eq36}
 x_{i,\, k} = S_{i,\, k} + z_{i,\, k}
\end{equation}
where $z_{i,\, k}$ denotes real-valued samples of AWGN with variance
$\sigma_w^2/2$ \cite{Vasu_Book10}. We assume that both $p(\cdot)$ and
$\hat{p}_{\mathrm{m}}(\cdot)$ have unit energy, that is
\begin{align}
\label{Eq:FLOGEN_Eq37}
\sum_{m=-M}^{M}
 p^2(mT_s)                      & = 1                            \nonumber  \\
\sum_{m=-M}^{M}
\hat{p}^2_{\mathrm{m}}(mT_s)    & = 1.
\end{align}
Note that from (\ref{Eq:FLOGEN_Eq10}), (\ref{Eq:FLOGEN_Eq11}) and
(\ref{Eq:FLOGEN_Eq14}), $x_{i,\, k,\, I},\, x_{i,\, k,\, Q}$ in
Figure~\ref{Fig:FLOGEN_Dis_Sys_Model} have to be summed (or averaged). Since
$S_{i,\, k}$ carries half a bit of information, the
average signal-to-noise ratio (SNR) per bit is defined as \cite{Vasu_Book10}
\begin{align}
\label{Eq:FLOGEN_Eq38}
\mathrm{SNR}_{\mathrm{av},\, b} 
& = \frac{2E\left[S_{i,\, k}^2\right]}
         {\mbox{2D noise variance}}                 \nonumber  \\
& = \frac{2\times 1}{2\times \sigma^2_w/2}                  \nonumber  \\
& = \frac{2}{\sigma^2_w}.
\end{align}
\section*{\centering\normalsize\bf SIMULATION RESULTS}
\label{Sec:Sim_Results}
\begin{table}[tbhp]
\begin{center}
\caption{Simulation parameters.}
\input{flogen_sim_param.pstex_t}
\label{Tbl:FLOGEN_Sim_Param}
\end{center}
\end{table}
The discrete-time simulation parameters are given in
Table~\ref{Tbl:FLOGEN_Sim_Param}. The transmit filters are given by
$p(mT_s-0.5T_s),\, \hat{p}_{\mathrm{m}}(mT_s-0.5T_s)$, for $-M\le m\le M$,
for integer $m$ and $M=8N,\, 16N$. This is because, we require the filter
length to be an integer multiple of $N$, for the sake of implementation
simplicity.
\begin{figure}[tbhp]
\begin{center}
\input{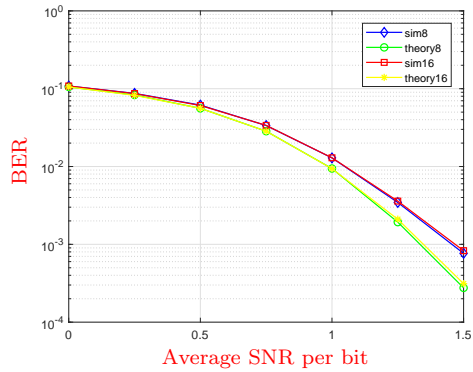}
\caption{Simulation results.}
\label{Fig:FLOGEN_BER}
\end{center}
\end{figure}
The computer simulation results for the bit-error-rate (BER) vs the
average SNR per bit ($\mathrm{SNR}_{\mathrm{av},\, b}$)
are presented in Figure~\ref{Fig:FLOGEN_BER}.
The theoretical BER is obtained from \cite{Vasudevan23}. The BER results
for $M=8N$ are denoted by ``{\tt sim8}'' and  ``{\tt theory8}''. The BER results
for $M=16N$ are denoted by ``{\tt sim16}'' and ``{\tt theory16}''. Observe the
close match between theory and simulations.
\section*{\centering\normalsize\bf CONCLUSIONS}
\label{Sec:Conclude}
This article discusses the implementation of an OFDM-OQAM/FBMC/UFMC system
in discrete-time, using Hilbert transform. A simple matched filter
receiver is sufficient for detecting the symbols. In the present work, an
AWGN channel is considered. Frequency selective
fading channels along with carrier frequency offsets can be considered in
future works.
\bibliographystyle{IEEEtran}
\bibliography{mybib,mybib1,mybib2,mybib3,mybib4,mybib5}
\end{document}